\documentstyle[12pt]{article}

\begin{document}
\thispagestyle{empty}

\begin{center}
\LARGE \tt \bf{Spacetime defects :

Domain walls and torsion.}
\end{center}


\begin{center}
{\large L.C. Garcia de Andrade\footnote{Departamento de
F\'{\i}sica Te\'{o}rica - Instituto de F\'{\i}sica - UERJ

Rua S\~{a}o Fco. Xavier 524, Rio de Janeiro, RJ

Maracan\~{a}, CEP:20550-003 , Brasil.

E-mail : GARCIA@SYMBCOMP.UERJ.BR}}
\end{center}


\begin{abstract}
The theory of distributions in non-Riemannian spaces is used to obtain exact static thin domain wall solutions of Einstein-Cartan equations of gravity. Curvature $ \delta $-singularities are found while Cartan torsion is given by Heaviside functions. Weitzenb\"{o}ck planar walls are caracterized by torsion $\delta$-singularities and zero curvature. It is shown that Weitzenb\"{o}ck static thin domain walls do not exist exactly as in general relativity. The global structure of Weitzenb\"{o}ck nonstatic torsion walls is investigated. 
\end{abstract}


\begin{center}
\end{center}
\newpage
\pagestyle{myheadings}
\markright{\underline{Spacetime defects : Domain walls and torsion.}}
\section{Introduction}
\paragraph*{}

The geometry of $ \delta $-torsion singularities \cite{1} have been extensively studied in connection with spacetime defects \cite{2,3,4}. An interesting class of Weitzenb\"{o}ck spacetime defects representing torsion loops and line defects representing cosmic strings was recently investigated by Letelier \cite{5}. Besides, pure Riemannian defects (disclinations) with zero Cartan's torsion over lines (strings) and walls (domain walls) have been investigated by Letelier \cite{5,6} and Wang. In their study they have considered that metric is continuous across the junction and the first derivative is represented by a Heaviside function. Thus in their approach the $\delta$-singularity just appears on the Riemann curvature.  We shall adopt here the same procedure when dealing with Riemann Cartan domain walls.

A Riemann-Cartan domain wall solution of Einstein-Cartan field equation is found where the torsion 2-form is a Heaviside function and the curvature is of  $\delta$-singularity type like in Letelier and Wang \cite{5,6} Riemannian domain walls.The solution is obtained by assuming the symmetric energy-momentum tensor in Einstein-Cartan gravity is the same as the general relativistic thin domain wall tensor proposed by Vilenkin \cite{7}. In the case of static Weitzenb\"{o}ck wall it is shown that the Vilenkin thin domain wall energy-momentum tensor does not satisfy the Einstein-Cartan field equations. Therefore by gluing together pieces of Minkowski spaces is possible to find solutions of Einstein-Cartan field equations as is done in Einstein field equations of general relativity \cite{8}. Thick domain walls in Einstein-Cartan theory of gravitation can be investigated in near future. In a certain sense this paper extends the investigation carried out by A.Wang \cite{9,10} on  plane-symmetric walls in general relativity. The important features of this paper is the fact that both torsion and curvature are necessary to guarantee that static thin domain walls exist in Einstein-Cartan theory of gravity.

\section{Geometry of Weitzenb\"{o}ck walls}
\paragraph*{}

Spacetime defects \cite{2,5,6} appears in distinct forms. For example they appear as torsion line defects (cosmic strings) or as torsion loops. Domain walls appears in the context of general relativity as  another type of defects in spacetime. In particular plane cosmic walls \cite{11} have been studied in the case of plane gravitational collapse. I shall be concerned with the investigation of a type of defect which is very common in the differential geometry of fractures where torsion $\delta$-functions are generated from Heaviside functions. Here a static plane symmetric spacetime with zero Riemann-Cartan curvature and nonvanishing torsion, the so-called Weitzenb\"{o}ck spacetime; is found where one of the metric functions is a Heaviside function.

The metric is therefore piecewise constant and torsion is a $\delta$ - singularity which changes when one passes from one plane to the other. The Weitzenb\"{o}ck geometry of fractures is then developed. Let us now consider the plane symmetric metric

\begin{equation}
ds^{2}=({\omega}^{0})^{2}-({\omega}^{1})^{2}-({\omega}^{2})^{2}-({\omega}^{3})^{2}
\label{1}
\end{equation}
where
\begin{equation}
\begin{array}{llll}
{\omega}^{0} & = & e ^{\frac{F}{2}}dt \nonumber \\
{\omega}^{1} & = & e ^{\frac{H}{2}} dx \nonumber \\
{\omega}^{2} & = & e ^{\frac{H}{2}}dy \nonumber \\
{\omega}^{3} & = & e ^{\frac{G}{2}} dz \nonumber
\end{array}
\label{2}
\end{equation}

The torsion 2-forms are chosen according to 

\begin{equation}
\begin{array}{llll}
T^{0} & = & J^{(0)} {\omega}^{0} \wedge {\omega}^{3} \nonumber \\
T^{1} & = & J^{(1)} {\omega}^{3} \wedge {\omega}^{1} + J^{(2)} {\omega}^{0} \wedge {\omega}^{1} \nonumber  \\
T^{2} & = & J^{(1)} {\omega}^{3} \wedge {\omega}^{2} + J^{(2)} {\omega}^{0} \wedge {\omega}^{2} \nonumber \\
T^{3} & = & J^{(3)} {\omega}^{0} \wedge {\omega}^{3} \nonumber
\end{array}
\label{3}
\end{equation}

Making use of Cartan's calculus where the torsion 2-forms is given by

\begin{equation}
T^{a}=d{\omega}^{a}+{{\omega}^{a}}_{b} \wedge {\omega}^{b} \mbox{(a,b=0,1,2,3)}
\label {4}
\end{equation}
one obtains the following 1-form connections

\begin{equation}
\begin{array}{llll}
{{\omega}^{0}}_{3} & = & \lbrack J^{(0)} + e^{\frac{-G}{2}} \frac{F'}{2} \rbrack {\omega}^{0} \nonumber \\
{{\omega}^{1}}_{0} & = & \lbrack J^{(2)} + e^{\frac{-F}{2}} \frac{\dot{H}}{2} \rbrack {\omega}^{0} \nonumber \\
{{\omega}^{1}}_{3} & = & \lbrack J^{(1)} + e^{\frac{-G}{2}} \frac{H'}{2} \rbrack {\omega}^{1} \nonumber \\
{{\omega}^{2}}_{3} & = & - \lbrack J^{(1)} + e^{\frac{-G}{2}} \frac{H'}{2} \rbrack {\omega}^{2} \nonumber
\end{array}
\label{5}
\end{equation}
and $ J^{(3)} = \frac{\dot{G}}{2} e^{\frac{-F}{2}} $

Making the choices

\begin{eqnarray}
J^{(0)} & = & - e^{\frac{-G}{2}} \frac{F'}{2} \nonumber \\
J^{(1)} & = &  e^{\frac{-G}{2}} \frac{H'}{2} \\
J^{(2)} & = &  e^{\frac{-F}{2}} \frac{\dot{H}}{2} \nonumber
\label{6}
\end{eqnarray}
one notices that for this choice of torsion functions $ {\omega}^{a}_{b} \equiv 0 $ and from the second Cartan's structure equation

\begin{equation}
{{R}^{a}}_{b}= d{{\omega}^{a}}_{b} +  {{\omega}^{a}}_{b} \wedge {{\omega}^{c}}_{b}
\label{7}
\end{equation}
${R^{a}}_{b}=0$ where $ {{R}^{a}}_{b} \equiv {{R}^{a}}_{bcd} {\omega}^{c} \wedge {\omega}^{d} $ are the curvature 2-forms, thus the curvature $ {{R}^{a}}_{bcd} $ vanishes.

Let us now consider the static case where $ \dot{H}=\dot{G}=\dot{F}=0$.

To simplify matters one considers $ J^{(0)} \equiv 0 $ ,which implies $  F'=0 $ and $  F=const.$ . Making the choice $ G=G(z)=const. $ the only surviving equation is 

\begin{equation}
H'(z)= c_{1} J^{(1)}
\label{8}
\end{equation}
where $ c_{1} $ is a constant. Choosing $ J^{(1)} \equiv \delta (z) $ where $ {\delta}(z)$ is Dirac ${\delta}$-function, equation (\ref{8}) has the following solution

\begin{equation}
H(z) \equiv {\theta}_{0}(z)=
\left \{
\begin{array}{ll}  
1  , z > 0 \nonumber \\
0  , z < 0 \nonumber \\
\end{array}
\right.
\end{equation}
where ${\theta}_{0}(z)$ is the Heaviside function. Thus the metric is basically given by a Heaviside function and torsion is a $ {\delta} $ -singularity \cite{6}.

Notice that outside the defect ($ z \neq  0 $)$ H(z)=0 $,$ F=0 $, $ G=\mbox{const.}$ and the spacetime is Minkowskian. $ ds^{2} = dt^{2} - dx^{2}- dy^{2}-dz'^{2} $ where  $ dz'^{2}=c^{2} dz^{2} $ and since c is a constant $ z'= c z $ which means a translation along the z-axis. To resume we show that a new class of fractures in Weitzenb\"{o}ck spacetime exists besides torsion loops and torsion lines.

From this computation it is easy to show that static thin  Weitzenb\"{o}ck domain walls are forbidden in Einstein-Cartan theory of gravity. Thin domain walls have the following energy-momentum tensor

\begin{equation}
{{T}_{a}}^{b}= {\sigma} diag(1,1,1,0) {\delta}(z)
\label{10}
\end{equation}

From the Einstein-Cartan field equations
\begin{equation}
R_{ab}({\Gamma})- \frac{1}{2} g_{ab} R({\Gamma})=T_{ab}
\label{11}
\end{equation}
and the Weitzenb\"{o}ck condition of vanishing total curvature $ R_{abcd}({\Gamma})=0 $ it is obvious that $ {\sigma}=0 $ and thus there is no static thin domain wall formation in Weitzenb\"{o}ck universe. Therefore we are compelled to consider Riemann-Cartan domain walls. In fact in the next section I shall present a simple solution of Einstein-Cartan field equations of gravity representing a static thin domain walls.

\section{Riemann-Cartan thin domain walls}
\paragraph*{}

In this section I shall be concerned with finding a solution to Einstein-Cartan theory of gravity which describes a thin domain wall \cite{12} given by the energy-momentum tensor equation (\ref{10}).

I shall compute this solution by solving the system composed by the first and second Cartan's equations of structure given in the last section and the Einstein-Cartan field equation (\ref{11}). From the 1-form connections (\ref{6}) and the second Cartan structure equation after some algebra one obtains the components of the Riemann-Cartan non-vanishing components of curvature

\newpage

\begin{equation}
\begin{array}{llll}
{{R}^{2}}_{112} & = & - \frac{(H')^{2}}{4} e^{-F}\nonumber \\
{{R}^{2}}_{332} & = & - \frac{e^{-F}}{2} \lbrack H'' + \frac{H'F'}{2} - \frac{(H')^{2}}{2} \rbrack = {{R}^{3}}_{223}\nonumber \\
{{R}^{1}}_{331} & = & - \frac{e^{-F}}{2} \lbrack H'' + \frac{H'F'}{2} \rbrack \nonumber \\
{{R}^{0}}_{330} & = & - \frac{e^{-F}}{2} \lbrack 2 {{J}^{0}}^{'} - \frac{F'{J}^{0}}{2} + \frac{{F'}^{2}}{2} \rbrack \nonumber \\
{{R}^{0}}_{110} & = &  {{R}^{2}}_{002} = -\frac{e^{-F}}{2} \lbrack 2 {{J}^{0}}^{'} + \frac{F'H'}{2} \rbrack \nonumber
\end{array}
\label{12}
\end{equation}
where to simplify matters we chose $ J^{0} $ as the only nonvanishing component of Cartan's torsion 2-forms. This choice is compatible with Cartan's first equation of structure (\ref{4}). The metric is also chosen to be static and Vilenkin hypotesis $ F \equiv G $ is also used. The Einstein tensor is defined as

\begin{equation}
G_{ab} = R_{ab}({\Gamma}) - \frac{1}{2} g_{ab}R({\Gamma})
\label{13}
\end{equation}
Where $ {\Gamma} $ is the Riemann-Cartan connection.

The Ricci-Cartan tensor and Ricci scalar are given by

\begin{eqnarray}
R_{00} & = & \frac{e^{-F}}{2} \lbrack \frac{F'}{2} J^{0} - \frac{{F'}^{2}}{2} \rbrack \nonumber \\
R_{11} & = & \frac{e^{-F}}{2} \lbrack 2 {{J}^{0}}^{'}  + \frac{{H'}^{2}}{2} + H'' + F'H' \rbrack \\
R_{22} & = & \frac{e^{-F}}{2} \lbrack - 2 {{J}^{0}}^{'} + H'' \rbrack \nonumber  \\
R_{33} & = & \frac{e^{-F}}{2} \lbrack 2 {{J}^{0}}^{'}  - \frac{F'{J}^{0}}{2} + \frac{{F'}^{2}-{H'}^{2}}{2} - H'F' \rbrack \nonumber
\label{14}
\end{eqnarray}
and

\begin{equation}
R \equiv {{R}^{0}}_{0} + {{R}^{1}}_{1} + {{R}^{2}}_{2} + {{R}^{3}}_{3} =\frac{e^{-F}}{2} \lbrack H''+ F'H'-2{{J}^{0}}^{'} + F'J^{0} - {F'}^{2} \rbrack
\label {15}
\end{equation}

Collecting all those terms, substitution into the Einstein-Cartan eqns. yields
\begin{equation}
\begin{array}{llll}
G_{00} & = & \frac{e^{-F}}{2} \lbrack 2 {{J}^{0}}^{'}  - H'' - \frac{F'H'}{2} \rbrack = {\sigma}{\delta}(z) \nonumber \\
G_{11} & = & \frac{e^{-F}}{2} \lbrack H'' - \frac{{H'}^{2}}{2} + \frac{{F'H'}^{2}}{2} \rbrack = -  {\sigma}{\delta}(z) \nonumber \\
G_{22} & = & \frac{e^{-F}}{2} \lbrack -4 {{J}^{0}}^{'} + H'' + \frac{H'F'}{2} + \frac{F'{J}^{0}}{2} - {F'}^{2} \rbrack = - {\sigma}{\delta}(z) \nonumber \\
G_{33} & = & \frac{e^{-F}}{2} \lbrack - \frac{{H'}^{2}}{2} + H'' - \frac{F'H'}{2} \rbrack = 0 \nonumber
\end{array}
\label{16}
\end{equation}

This system of nonlinear ODE can be easily solved if one consider $ F'=0 $. With this choice Cartan torsion reduces to a Heaviside function since

\begin{equation}
{{J}^{0}}^{'}=\frac{c{\sigma}}{4}{\delta}(z) \hspace{.5cm}  \rightarrow \hspace{.5cm} {J}^{0}=\frac{c{\sigma}}{2}{\theta}_{0}(z)
\label{17}
\end{equation}

And the metric function H'' reads
\begin{equation}
H'' = - \frac{c{\sigma}}{2}{\delta}(z) \hspace{.5cm}  \rightarrow  \hspace{.5cm} H'' = - \frac{c{\sigma}}{2}{\theta}_{0}(z)
\label{18}
\end{equation}

Expression (\ref{17}) means that torsion undergoes a jump across the hyperplane $ z=0 $ while from (\ref{18}) most of the curvature components behaves like a $ \delta$-singularity. For example $ {{R}^{0}}_{330}={{R}^{0}}_{220}={{R}^{0}}_{110}={{R}^{1}}_{331}={{R}^{2}}_{332} =c {\sigma}{\delta}(z) $ while $ {{R}^{2}}_{112}= -\frac{{c}^{2}}{4}{\sigma}^{2}{{\theta}_{0}}^{2}(z) $ which is the square of the Heaviside function. 

From (\ref{18}) expression (\ref{12}) reduces to 
\begin{eqnarray}
{{R}^{0}}_{330} & = &  {\tau}[ 2 {\sigma}{\delta}(z) ] \nonumber \\
{{R}^{0}}_{110} & = &  {{R}^{2}}_{002} = {\delta}(z)k \nonumber \\
{{R}^{1}}_{331} & = &  c[{\sigma}{\delta}(z)] \\
{{R}^{2}}_{112} & = &  c{{H'}^{2}} = - \frac{c^{2}}{4} {\sigma}^{2} {\theta}_{0}(z) \nonumber \\
{{R}^{2}}_{332} & = &  c[ 2 {\sigma}{\delta}(z)] \nonumber
\label{19}
\end{eqnarray}
where $ k \equiv 2c  $. All curvature components vanishes but $ {{R}^{2}}_{112} $  and  $ {{R}^{2}}_{332}$ . Since the spacetime outside z=0 plane has to be locally flat vanishing of these components leads to $  H'=0 $ which yields $ H=const. $ and the spacetime metric is Minkowskian. Thus our domain wall geometry is composed of two Minkowski half-spaces  glued together across a torsion and curvature junction.  As in Letelier \cite{5} paper no attempt is made to find a field theory interpretation for the solution discussed here. It is easy to show that the Bianchi identities are identically satisfied.

This solution is similar to a solution found by Holvorcen and Letelier \cite{12} for plane symmetric cosmic walls and gravitational collapse. To resume we have investigated two classes of solutions that represent static thin planar walls in spacetime with torsion. In one of the solutions torsion is a $ {\delta} $-singularity. In the other torsion is a Heaviside function while the Riemannian curvature posseses a $ {\delta}$-singular behaviour. Notice that no physical interpretation is tried for the Weitzenb\"{o}ck class of solutions. Nevertheless they may be represent a class of gravitational ghosts like ghost neutrinos where the stress-energy tensor vanishes although the current is distinct from zero. Indeed our solution is not able to represent ghost neutrinos since our axial torsion vanishes and thus no axial current of ghost neutrinos \cite{13} is possible. Nevetheless Letelier has found recently \cite{2} a class of Weitzenb\"{o}ck spacetime defects where an axial torsion exists, being a candidate for a ghost neutrino space. Torsion and curvature $ {\delta} $-singularities have also been found in Jackiw \cite{14} three dimensional planar gravity.
Since domain walls in the large scale structure of the Universe are either unstable or thick, the domain wall solution discussed here can only be considered as a  thin domain wall approximation of a static thick domain wall in Riemann-Cartan spacetime. Work in this direction shall be addressed in near future.  A detailed investigation of the global properties of Riemann-Cartan domain walls can appear elsewhere.

\section{Global Structure of torsion walls}
\paragraph*{}

The investigation of maximal extension of metrics in Riemann-Cartan Space-time like the one studied in the last section can be extremely difficult.  Therefore to simplify matters in this section I shall address some important questions concerning nonstatic Weitzenb\"{o}ck torsion wall which answers many questions which appears in the more general case of domain walls in Riemann-Cartan spaces. Therefore let us consider the metric used by  \cite{9}  to investigated gravitationally repulsive domain wall solutions of Einstein's equations of general relativity. The metric is given by

\begin{equation}
ds^{2}= e^{2{\nu}(t,\vert z \vert)} (-dt^{2} + dz^{2}) + B(t,\vert z \vert)(dx^{2}+dy^{2})
\label{20}
\end{equation}
in the notation of Cartan's calculus metric (\ref{20}) reads

\begin{equation}
ds^{2}=-({\omega}^{0})^{2} + ({\omega}^{1})^{2} + ({\omega}^{2})^{2} + ({\omega}^{3})^{2}
\label{21}
\end{equation}

Where the basis 1-forms $ {\omega}^{a} \hspace{0.3cm} (a=0,1,2,3) $ are
\begin{equation}
\begin{array}{llll}
{\omega}^{0} & = & e^{\nu} dt \nonumber \\
{\omega}^{1} & = & e^{\nu} dz \nonumber \\
{\omega}^{2} & = & \sqrt{B} dx \nonumber \\
{\omega}^{3} & = & \sqrt{B} dy\nonumber \\
\end{array}
\label{22}
\end{equation}

Choosing Cartan torsion 2-forms a priori by 
\begin{equation}
T^{A}=J^{A}{\omega}^{0} {\wedge} {\omega}^{1}
\label{23}
\end{equation}
where (A=0,1) and $ T^{i}=0 $ for (i=2,3) and using Cartan's first structure equation (\ref{4}) one is able to compute the following  connection 1-forms

\begin{equation}
\begin{array}{lllll}
{{\omega}^{0}}_{1} = (J^{0}+ e^{-{\nu}}{\nu}_{z}){\omega}^{0} \nonumber \\
{{\omega}^{2}}_{0} = \frac{1}{2} \frac{B_{t}}{B} e^{-\nu}{\omega}^{2} \nonumber \\
{{\omega}^{3}}_{0} = \frac{1}{2} \frac{B_{t}}{B} e^{-\nu}{\omega}^{3} \nonumber \\
{{\omega}^{2}}_{1} = \frac{1}{2} \frac{B_{z}}{B} e^{-\nu}{\omega}^{2} \nonumber \\
{{\omega}^{3}}_{1} = \frac{1}{2} \frac{B_{z}}{B} e^{-\nu}{\omega}^{3} \nonumber \\
\end{array}
\label{24}
\end{equation}
and

\begin{equation}
J^{1}=e^{-{\nu}}{\nu}_{t}
\label{25}
\end{equation}
the others zero. From the second Cartan's structure equation (\ref{7})

One notices that the Weitzenb\"{o}ck condition of zero curvature can be obtained by making $ {{\omega}^{a}}_{b} \equiv 0 $. Substitution of these expression into (\ref{24}) yields

\begin{equation}
\begin{array}{ll}
J^{0} = -e^{-{\nu}}{\nu}_{z} \nonumber \\
B_{t} = B_{z}  =  0 \nonumber \\
\end{array}
\label{26}
\end{equation}
Since by definition a Wietzenb\"{o}ck torsion wall is one where the torsion must be equal to a ${\delta}$-singularity one should impose this condition on expression (\ref{26}).  This procedure can be obtained by considering the following ansatz 

\begin{equation}
{\nu} = - (t - {\theta}_{0}(z))
\label{27}
\end{equation}

Substitution of (\ref{27}) into (\ref{27}) yields immediatly

\begin{equation}
\begin{array}{ll}
J^{0}= - e^{(t-{\theta}_{0}(z))} {\delta}(z) \\
J^{1}= - e^{(t-{\theta}_{0}(z))} \\
\end{array}
\label{28}
\end{equation}
for $ \delta $-singularity distributions of torsion forms. From the second eqn. in (\ref{26}) one obtains $ B=B_{0}=const. $.  Thus the non-static spacetime metric is 

\begin{equation}
ds^{2}= e^{-2 (t-{\theta}_{0}(z))} (-dt^{2} + dz^{2}) + {(B_{0})}^{2}(dx^{2}+dy^{2})
\label{29}
\end{equation}

Metric (\ref{29}) can be rewriten in a conformal form 

\begin{equation}
ds^{2}= e^{-2(t-{\theta}_{0}(z))} [ -dt^{2} + dz^{2} + {(B_{0})}^{2} e^{2(t-{\theta}_{0}(z))} (dx^{2}+dy^{2})]
\label{30}
\end{equation}

From the properties of $ {\theta}_{0} $ function above one has outside the torsion wall

\begin{equation}
ds^{2}= {\alpha}e^{-2t} [ -dt^{2} + dz^{2} + {(B_{0})}^{2} e^{2t} (dx^{2}+dy^{2})] \hspace{.5cm} (z<0)
\label{31}
\end{equation}
where $ {\alpha} $ is a constant.

This metric is nothing more than a metric conformal to de Sitter spacetime.

At $ z>0 $ the spacetime is

\begin{equation}
ds^{2}= e^{-2 (t-1)} (-dt^{2} + dz^{2}) +{(B_{0})}^{2}(dx^{2}+dy^{2})
\label{32}
\end{equation}
or

\begin{equation}
ds^{2}= {\alpha}' e^{-2t} [ -dt^{2} + dz^{2} + {\gamma}' e^{2t} (dx^{2}+dy^{2})] \hspace{.5cm} (z>0)
\label{33}
\end{equation}

At t=0 the spacetimes outside the wall are flat as carachterized by a spacetime defect.  As $ t \to \infty $ from  (\ref{26}) the torsion $ J^{0} $ also "explodes" and the torsion wall suffers gravitational colapse. The solution at z=0 is obtained by gluing together the two half de Sitter spaces given by the line elements (\ref{31}) and (\ref{33}).

Let us now consider the motion eqns. In general when one deals with space time with torsion test particles follow to distinct paths. The first are autoparallels and the second are geodesics. Since we are just concerned with Weitzenb\"{o}ck spaces where the Riemann-Cartan connection vanishes autoparallel are straight lines. Thus geodesic eqns. reads

\begin{equation}
\begin{array}{lll}
\ddot{t} + \dot{t}^{2} + \dot{z}^{2} = 0 \\
\ddot{z} + \dot{t} \dot{z}=0 \\
\ddot{x} = \ddot{y} = 0 \\
\end{array}
\label{34}
\end{equation}
where dots in (\ref{34}) mean derivation with respect to proper time S. The geodesic of the z-coordinate is easily solved by $ \dot{z}=e^{-t} $ and $\ddot{z}=-e^{-t} $. Thus considering $ \dot{t}>0  \vspace{.4cm}(s>0)$, $ \dot{z}>0 $ and $ \ddot{z}<0 $ and since the speed of the test particle vanishes at $ t \to \infty $ and the acceleration is opposite to the velocity the test particle falls back to the torsion defect. Test particle is clearly bound. Notice that another solution to the geodesic motion would be $ \dot{z}=- e^{-t} $ and $ \ddot{z}=e^{-t} \dot{t} $ and in this cases test particles are moving towards the torsion defect at $ z=0 $ plane. Since once more acceleration is opposite to the velocity it may be possible in principle that an repulsive gravitational force exists at the torsion wall and can be used to hold or at least decrease the speed of the gravitational collapse due to torsion effects. 

More realistic models of non static domain walls in Riemann-Cartan space-time may be found else where.  Discussion of stability of cosmic strings in general relativity may be found in the paper by R.Gleiser and J.Pullin \cite{15}.  The issue of stability of domain walls in the context of Einstein-Cartan gravity may be found elsewhere as well as the study of transparency of domain walls to neutrino waves.

This is closed related to A. Trautman \cite{16} idea to use spin and torsion in the context of Einstein-Cartan theory of gravity to hold or avert gravitational collapse of stellar objects. A spacetime defect to be pure \cite{5} needs to be empty of matter ad fields outside the defect. Therefore to obey the rule we must force the torsion 2-form component $ J^{1} $ to vanish. In this case from expressions (\ref{25}) $ {\nu}_{t} $ and the metric shall not depend on time coordinate . Thus a static torsion defect is obtained where Cartan torsion is given only by Dirac ${\delta}$-function.

From metric (\ref{29}) it is easily seen that the space-time off the wall is flat.  Therefore the metric off the wall is time independent.  However from expression (\ref{28}) it is seen that the thin wall is not static.  This situation happens also in the study of thin domain walls in general relativity.

\section*{Acknowledgements}
\paragraph*{}
I am very much indebt to Prof. P.S.Letelier and Prof. A.Wang for helpful discussions on the subject of this paper. Financial support from CNPq. and UERJ (FAPERJ) is gratefully acknowledged.

\end{document}